\documentclass[aps,twocolumn,showpacs,floatfix, pra]{revtex4-1}
\usepackage{hyperref,ulem}
\hypersetup{colorlinks=true, urlcolor=blue, citecolor=blue, linkcolor=blue}

\usepackage{epsfig}
\usepackage{newlfont}
\usepackage{amssymb}
\usepackage{amsfonts}
\usepackage{amsmath}
\usepackage{bm}
\usepackage{epstopdf}
\usepackage{graphicx}

\def\lsim{\mathrel{\rlap{\lower4pt\hbox{\hskip1pt$\sim$}}
    \raise1pt\hbox{$<$}}}                
\def\gsim{\mathrel{\rlap{\lower4pt\hbox{\hskip1pt$\sim$}}
    \raise1pt\hbox{$>$}}}                

\begin{document}

\title{Quantifying quantum correlation of quasi-Werner state and probing its suitability for quantum teleportation}
\author{Arpita Chatterjee\(^{1,}\)}\email{arpita.sps@gmail.com}
\author{Kishore Thapliyal\(^{2,}\)}\email{kishore.thapliyal@upol.cz}
\author{Anirban Pathak\(^{3,}\)}\email{anirban.pathak@jiit.ac.in}

\affiliation{\(^1\)J. C. Bose University of Science and Technology, YMCA, Faridabad 121006, Haryana, India}

\affiliation{\(^2\)Joint Laboratory of Optics of
Palack\'{y} University and Institute of Physics of CAS, RCPTM, Faculty of
Science, Palack\'{y} University, 17. listopadu 12, 771 46 Olomouc, Czech
Republic}

\affiliation{\(^3\)Jaypee Institute of Information Technology, A-10, Sector 62, Noida, UP 201309, India}

\date{\today}

\begin{abstract}
The significance of photon addition in engineering the single- and two-mode (bipartite correlations) nonclassical properties of a quantum state is investigated. Specifically, we analyzed the behavior of the Wigner function of two quasi-Werner states theoretically constructed by superposing two normalized bipartite $m$-photon added coherent states. This allowed us to quantify the amount of nonclassicality present in the quantum states using Wigner logarithmic negativity (WLN), while quantum correlations are measured in terms of concurrence, entanglement of formation, and quantum discord. The WLN for a two-mode state corresponds to the sum of the single-mode nonclassicality as well as quantum correlations, and both of these are observed to enhance with photon addition manifesting the efficacy of photon addition in the entanglement distillation. Usefulness of photon addition is further established by showing that the performance of the quasi-Werner states as quantum channel for the teleportation of a single-mode coherent and squeezed states, as quantified via teleportation fidelity, improves with the photon addition. Further, in contrast to a set of existing results, it is established that the negative values of two-mode Wigner function cannot be used in general as a witness of quantum correlation. 
\end{abstract}


\maketitle

\section{INTRODUCTION}

In 1935, the concept of quantum correlation, namely quantum entanglement, was introduced by Einstein \textit{et al.} \cite{einstein} and Schr\"{o}dinger \textit{et al.} \cite{schro}. Since then entangled states have been established as a promising candidate for many applications in quantum information technology \cite{tittel,knill,zoller}. {It is also found useful to improve our understanding of various foundational issues, ranging from the possibility of detecting graviton \cite{BH2} to the black hole information paradox \cite{BH1}.}  Moreover, the rapid development in quantum computation and communication has rendered further interest in the production and investigation of entangled resources. In such a context, a large number of theoretical as well as experimental schemes have been proposed
for generating the entangled states based on photonic \cite{kwiat,shih} and atomic \cite{hang,zheng} qubits as well as  qudit \cite{franco} and infinite dimensional \cite{howell,liao} states. With the advent of the newer applications of entanglement, it has become extremely important to quantify the amount of entanglement. As a consequence, several measures of entanglement  have been proposed. For example, concurrence \cite{bennet}, entanglement of formation \cite{rungta}, tangle \cite{coffman}, and negativity \cite{plenio} are some widely used measures for quantifying the entanglement exhibited in a bipartite or multipartite quantum state. 

Later, Ollivier and Zurek \cite{zurek} discovered that entanglement is not the only type of multiparty quantum correlation, and consequently the entanglement measures, like concurrence or entanglement of formation, cannot be considered as a complete measure of quantum correlation. They proposed quantum discord  (QD) \cite{zurek1}, a measure of quantum correlation, defined as the difference between the quantum versions of two classically equivalent expressions of mutual information. Since QD is based on the total correlation (mutual information), it can predict about quantum correlation in non-separable (entangled) as well as in separable
states \cite{mishra}. An appreciable amount of work related to the theoretical development of QD \cite{hender,oppen,vedral}, its dynamical property under the effect of decoherence \cite{maziero,welang}, and its comparison with entanglement \cite{yonac,bellomo} have been performed over the past two decades. The superiority of QD over entanglement in quantifying the quantum correlation and its applicability has kindled more interest in investigating the dynamics of QD in different quantum mechanical systems, like bipartite two-level atomic system interacting with a cavity field \cite{wang,yi}, quantum dots \cite{fanchini}, spin chains \cite{wer}, and in different quantum states, like bipartite Bell diagonal, bipartite X class \cite{mathh}, continuous variable Werner  \cite{CV-W}, non-Gaussian Werner \cite{bellomo1}, and multi-partite \cite{tathram} states. It may be apt to note that a Werner state \cite{werner} which can be viewed as a statistical mixture of a maximally entangled pure and a maximally mixed states, is often used in the studies of quantum correlations as well as coherence \cite{SMcoh}.  This motivated us to use quasi-Werner states to investigate the significance of photon addition (a local non-Gaussian operation) as a quantum state engineering tool to enhance the single- and two-mode (correlations) nonclassical properties of a quantum state. As far as photon addition is concerned, it is already established as a powerful tool for inducing and/or enhancing nonclassicality in single-mode Gaussian and non-Gaussian states (\cite{kishore1} and references therein). Some of these quantum engineered states were used recently for developing quantum technology, for instance, quantum metrology \cite{PM} and cryptography \cite{SS}.

 A variety of approaches have been proposed for the generation of nonclassical states in different optical systems, nonlinear optical processes, and cavity QED (\cite{co,PT} and references therein). Such type of radiation field not only provides a platform for testing fundamental concepts of quantum theory, but also for successful implementations of original quantum tasks, such as quantum teleportation, quantum cryptography (\cite{swit} and references therein). Interestingly, a state with the Wigner function \cite{wigner} taking negative values over some region of the phase space is nonclassical \cite{sadeghi} as they have non-positive Glauber-Sudarshan $P$ function which means that the quantum state cannot be represented as a statistical mixture of coherent states. However, the converse is not necessarily true, i.e., there exist states with positive Wigner function which show nonclassical properties, such as squeezed state.

In the recent years, a considerable amount of attention has been devoted to understand the connection between the negative Wigner values and other measures of quantum correlation. It has been
diagnosed that nonclassicality can be used as resource of a striking quantum feature, entanglement \cite{asboth,vanlook,paris}. A beam splitter is capable of converting nonclassicality of a single mode radiation into bipartite \cite{asboth} and multipartite \cite{zubairy20} entanglement.  The quantum correlations of two-mode continuous variable separable states are studied recently in \cite{taghia}, which reported two well-defined measures of quantum correlation, namely QD and local quantum
uncertainty. Further, a comparative study of QD and entanglement in quasi-Werner states based on bipartite entangled coherent states $\left(|\alpha, \alpha\rangle \pm |-\alpha, -\alpha\rangle\right)$ and $\left(|\alpha, -\alpha\rangle\pm|-\alpha, \alpha\rangle\right)$ is also reported \cite{ajay}. Such states are found useful in the teleporation of coherent states (see \cite{mitali17} for a review). Siyouri \textit{et al.} addressed the pertinence and efficiency of the Wigner function in detecting the presence of quantum correlations \cite{siyouri}. They considered quantum systems described by Werner states of a superposition of bipartite coherent states and showed that the Wigner function is not sensitive to any kind of quantum correlations except entanglement. 
Here, we set ourselves a task to quantify the role of photon addition in altering quantum correlations present in the quasi-Werner states and in the teleportation of single-mode coherent and squeezed states using them as quantum channels. During this attempt, we also critically analyze the claims made by Siyouri \textit{et al.} \cite{siyouri} and establish that the Wigner function of the quasi-Werner state remains negative even in the absence of entanglement unlike claimed by them.

The rest of the paper is structured as follows: Sec.~\ref{sec2} describes the quantum states of our interest. We have shed some light onto the nature of Wigner distribution and Wigner logarithmic negativity in next two sections. In Sec.~\ref{sec5}, we have compared the behavior of Wigner function and quantum correlations. The performance of the quasi-Werner states in teleporting a single-mode coherent as well as squeezed states are analyzed in Sec.~\ref{sec6} with specific attention to the role of the photon addition. The paper is concluded in Sec.~\ref{sec:Con}.

\section{States of interest}
\label{sec2}

An appropriate basis for describing many continuous variable quantum systems is a set composed of the so-called coherent states \cite{glauber63}. These states can easily be generated by applying a unitary operation, i.e., displacement operator $D(\alpha)$, to the vacuum state $|0\rangle$ of the quantized field as $|\alpha\rangle = D(\alpha)|0\rangle$. The Fock state representation of a single-mode coherent state is $|\alpha\rangle = e^{-\frac{1}{2}|\alpha|^2}\sum_{n=0}^\infty\frac{{\alpha}^n}{\sqrt{n!}}|n\rangle$. A photon-added coherent state is obtained by performing a non-unitary operation, creation operator ${a^\dagger}$, on the coherent state $|\alpha\rangle$. An $m$-photon-added coherent state is defined as \cite{agarwal91}
\begin{eqnarray*}
|\alpha, m\rangle & = & \frac{{a^\dagger}^m|\alpha\rangle}{\langle\alpha|a^m{a^\dagger}^m|\alpha\rangle^{1/2}}= \frac{{a^\dagger}^m|\alpha\rangle}{\big[m!\,L_m(-|\alpha|^2)\big]^{1/2}},
\end{eqnarray*}
where $m$ is an integer and $L_m(z)$ is the usual Laguerre polynomial of order $m$ \cite{ryzhik}.

If the coherent states $|\alpha\rangle$ and $|\beta\rangle$ are employed on the Hilbert spaces $\mathcal{H}^A$ and $\mathcal{H}^B$, respectively, then the two-mode state $|\alpha,\,\beta\rangle$ is represented in the tensor product space $\mathcal{H}^A\otimes\mathcal{H}^B$. Extending the idea, an $m$-photon-added bipartite coherent state may be defined as
\begin{eqnarray*}
|\alpha, \beta, m\rangle = \frac{{a^\dagger}^m {b^\dagger}^m|\alpha, \beta\rangle}{\langle\alpha, \beta|a^m b^m {a^\dagger}^m {b^\dagger}^m|\alpha, \beta\rangle^{1/2}},
\end{eqnarray*}
where $|\alpha\rangle$ and $|\beta\rangle$ are any two coherent states having amplitudes $\alpha$ and $\beta$, respectively. The states $|-\alpha\rangle$ and $|-\beta\rangle$ are $\pi$ radian out of phase with the corresponding coherent states $|\alpha\rangle$ and $|\beta\rangle$, respectively.

In this paper, we consider two superposed $m$-photon-added bipartite coherent states as
\begin{eqnarray}
|\psi^+\rangle & = & N_+\big[|\alpha, \beta, m\rangle + |-\alpha, -\beta, m\rangle\big],\nonumber\\
|\psi^-\rangle & = & N_-\big[|\alpha, \beta, m\rangle - |-\alpha, -\beta, m\rangle\big],
\label{eq1}
\end{eqnarray}
where the normalization factor $N_{\pm}$ can be computed as
\begin{eqnarray*}
N_{\pm} & = & \left\{\frac{e^{|\alpha|^2+|\beta|^2}L_m(-|\alpha|^2)L_m(-|\beta|^2)}{2L_m^{\pm}(|\alpha|^2, |\beta|^2)}\right\}^{1/2}
\end{eqnarray*}
with
\begin{eqnarray*}
L_m^{\pm}(x,y) = e^{x+y}L_m(-x)L_m(-y)
\pm e^{-x-y}L_m(x)L_m(y).
\end{eqnarray*}

Let us consider another basis formed by the $m$-photon-added even and odd coherent states in the following manner:
\begin{eqnarray}
\label{eq9}
|+_\alpha\rangle & = & n_+^\alpha\big[|\alpha, m\rangle+|-\alpha, m\rangle\big],\nonumber\\
|-_\alpha\rangle & = & n_-^\alpha\big[|\alpha, m\rangle-|-\alpha, m\rangle\big],\nonumber\\
|+_\beta\rangle & = & n_+^\beta\big[|\beta, m\rangle+|-\beta, m\rangle],\nonumber\\
|-_\beta\rangle & = & n_-^\beta[|\beta, m\rangle-|-\beta, m\rangle],
\end{eqnarray}
where the normalization constants are given by
\begin{eqnarray*}
n_\pm^\xi& = & \left[\frac{e^{|\xi|^2}L_m(-|\xi|^2)}{2\left\{e^{|\xi|^2} L_m(-|\xi|^2) \pm e^{-|\xi|^2} L_m(|\xi|^2)\right\}}\right]^{1/2}
\end{eqnarray*}
with $\xi\in \{\alpha,\beta\}$.
Then the superposed $m$-photon-added bipartite coherent states (\ref{eq1}) can be rewritten in terms of the basis (\ref{eq9}) as
\begin{eqnarray}\label{eqQB}
|\psi^+\rangle & = & \frac{N_+}{2}\bigg[\frac{|+_\alpha, +_\beta\rangle}{n_+^\alpha n_+^\beta}
+\frac{|-_\alpha, -_\beta\rangle}{n_-^\alpha n_-^\beta}\bigg],\nonumber\\
|\psi^-\rangle & = & \frac{N_-}{2}\bigg[\frac{|+_\alpha, -_\beta\rangle}{n_+^\alpha n_-^\beta}
+\frac{|-_\alpha, +_\beta\rangle}{n_-^\alpha n_+^\beta}\bigg].
\end{eqnarray}

Based on these superposed $m$-photon-added bipartite coherent states, two quasi-Werner states \cite{werner} are defined as
\begin{eqnarray}
\rho(\psi^{\pm}, a) & = & (1-a)\frac{I}{4}+a|\psi^\pm\rangle\langle\psi^\pm|,
\label{eq_st}
\end{eqnarray}
with $a$ being the mixing parameter ranging from 0 to 1 and $I$ as an identity matrix corresponding to maximally mixed state. Note that quasi-Werner states are different from continuous variable Werner  state in \cite{CV-W}. It is clear that the quantum correlations present in these quasi-Werner states depend on the mixing parameter $a$, the coherent state amplitudes $\alpha$, and $\beta$, as well as the photon excitation number $m$.

\section{Wigner distribution}
\label{sec3}

The nonclassicality of a quantum state $\rho$ can be studied well in terms of its phase-space distribution characterized by the Wigner function \cite{scully}. For instance, the Wigner function $W(z, z^*)$ corresponding to an $m$-photon-added coherent state $|\alpha, m\rangle$ can be evaluated in terms of the coherent state basis as \cite{arpita1,agarwal91}
\begin{eqnarray*}
W(z, z^*) & = & \frac{2}{\pi^2}e^{2|z|^2}\displaystyle \int d^2\beta \langle -\beta|\alpha, m\rangle\langle\alpha, m|\beta\rangle e^{2(\beta^* z-\beta z^*)}.
\end{eqnarray*}
On simplification, the Wigner function for the state $|\alpha, m\rangle$ reduces to
\begin{eqnarray}
\label{eq2}
W_{\alpha, \alpha} = \frac{2}{\pi}\frac{(-1)^mL_m(|2z-\alpha|^2)}{L_m(-|\alpha|^2)}\,\,e^{-2|z-\alpha|^2},
\end{eqnarray}
where, for simplicity, we have not written phase space parameter $z$ in the argument.

The Wigner function for a superposition of single-mode coherent states, like $|\alpha\rangle+|-\alpha\rangle$, is given by \cite{gerry05, usha06}
\begin{eqnarray}
W(z, z^*) = N_{\mathrm{cs}}\left[W_{\alpha, \alpha}+W_{-\alpha, -\alpha}+W_{\alpha, -\alpha}+W_{-\alpha, \alpha}\right],\label{eqW}
\end{eqnarray}
where $N_{\mathrm{cs}}$ is the normalization constant. Note that the first two terms in Eq.~(\ref{eqW}) correspond to the individual coherent states in the superposition and the last two terms appear as interference terms. 
Each term in Eq.~(\ref{eqW}) can be obtained proceeding in a way similar to the approach used in deriving  Eq.~(\ref{eq2}), specifically, a bit of computation would yield
\begin{eqnarray}
\label{eq3}
W_{-\alpha, -\alpha} = \frac{2}{\pi}\frac{(-1)^mL_m(|2z+\alpha|^2)}{L_m(-|\alpha|^2)}\,\,e^{-2|z+\alpha|^2}
\end{eqnarray}
and the interference terms
\begin{eqnarray}
\label{eq5}
W_{\alpha, -\alpha} & = & \frac{2}{\pi}\frac{(-1)^mL_m\big((2z-\alpha)(2z^*+\alpha^*)\big)}{L_m(-|\alpha|^2)}\nonumber\\
& & \times e^{-2|z|^2-2z\alpha^*+2z^*\alpha}\nonumber\\
&= & W_{-\alpha, \alpha}^*.
\end{eqnarray}
Further generalization to obtain Wigner function for a two-mode quantum state is obvious and yields \cite{wu97,jiang05}
\begin{eqnarray}
\label{eq6}
W(z_1, z_1^*, z_2, z_2^*) = W(z_1, z_1^*)\,\,W(z_2, z_2^*).
\end{eqnarray}
Using Eqs.~(\ref{eq2})-(\ref{eq5}) in Eq. (\ref{eq6}), the Wigner functions for two quasi-Werner states $\rho(\psi^{\pm}, a)$ are obtained as

\begin{eqnarray}
W_\pm(z_1, z_2) & = & \frac{1-a}{4\pi^2}+\frac{2a}{\pi^2}\frac{e^{|\alpha|^2+|\beta|^2}}{L_m^\pm(|\alpha|^2, |\beta|^2)} \Bigg[
L_m(|2z_1-\alpha|^2)\nonumber\\
& &L_m(|2z_2-\beta|^2)e^{-2|z_1-\alpha|^2}e^{-2|z_2-\beta|^2} \nonumber\\
& & +L_m(|2z_1+\alpha|^2)L_m(|2z_2+\beta|^2)e^{-2|z_1+\alpha|^2}\nonumber\\
& &e^{-2|z_2+\beta|^2}\pm \left\{ L_m\big((2z_1-\alpha)(2z_1^*+\alpha^*)\big)\right.\nonumber\\
& &L_m\big((2z_2-\beta)(2z_2^*+\beta^*)\big) e^{-2|z_1|^2-2|z_2|^2} \nonumber\\
& &\left.e^{-2z_1\alpha^*+2z_1^*\alpha}e^{-2z_2\beta^*+2z_2^*\beta}+{\textrm{c.c.}} \right\} \Bigg].
\label{eq7}
\end{eqnarray}
Here, ${\textrm{c.c.}}$ corresponds to the complex conjugate of the rest of the quantity in the curly brackets.

\begin{figure}[tbp]
\centering

\includegraphics[width=0.47\textwidth]{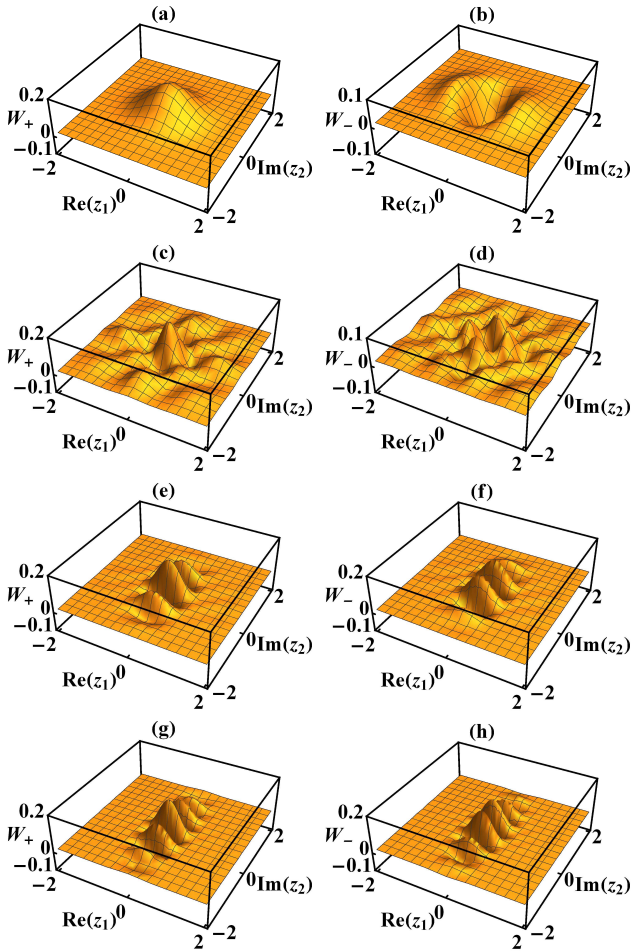} 

\caption{(Color online) Wigner function of the state $\rho(\psi^+, a)$ in Column 1 and $\rho(\psi^-, a)$ in Column 2 with (a)-(b) $\alpha = 0.2$, $\beta=0.1$, $a=0.4$, $m=0$, (c)-(d) $\alpha = 0.2$, $\beta=0.1$, $a=0.4$, $m=2$, (e)-(f) $\alpha = 1.2$, $\beta=1.1$, $a=0.4$, $m=2$,
 (g)-(h)  $\alpha = 1.2$, $\beta=1.1$, $a=0.4$, $m=4$, respectively.}

\label{fig1}
\end{figure}

Using Eq.\,(\ref{eq7}), the Wigner functions of quasi-Werner states are illustrated in Fig.~\ref{fig1} for different values of $\alpha$, $\beta$, $a$ and $m$ in the phase-space. It is easy to observe that if no photon is added at the beginning, $\rho(\psi^+,\,a)$ shows a Gaussian peak while $\rho(\psi^-,\,a)$ has a crater at the center [cf. Fig.~\ref{fig1}{(a)-(b)}]. This implies that for $m=0$, $\alpha=0.2$ and $\beta=0.1$, the Wigner function reveals the nonclassical character of $\rho(\psi^-,\,a)$, but cannot provide a conclusive witness of nonclassicality for $\rho(\psi^+,\,a)$ because {negativity of the Wigner function is a clear signature of nonclassicality of the related state}, but it is a one-sided condition. If consider $\alpha$, $\beta$ constant and raise $m$ from 0 to 2, a number of peaks and troughs of low height are found to be elevated in the surroundings of the central Gaussian peak of $\rho(\psi^+,\,a)$ [cf. Fig.~\ref{fig1}{(c)}]. This can be attributed to the contribution from higher-orders of the Laguerre polynomial as its zeroth order is unity. In a similar way, for $\rho(\psi^-,\,a)$, the crater in the middle is enclosed by a few small peaks as shown in Fig.~\ref{fig1}{(d)}. Furthermore, keeping $m$ fixed and increasing $\alpha$ and $\beta$ values, the contribution of interference terms dominates over the role of photon addition and thus distribution of $\rho(\psi^+,\,a)$ appears similar to that for the same $\alpha$ and $\beta$ without photon addition [see Fig.~\ref{fig1}{(e)}]. The number of interference patterns in the phase space further increases for the higher number of photons added [see Fig.~\ref{fig1}{(g)} for $m=4$].
The distribution of $\rho(\psi^-,\,a)$ also behaves similarly, but the behavior of $W_{-}(\textrm{Re}(z_1), \textrm{Im}(z_2))$ remained more relatable to $-W_{+}(\textrm{Re}(z_1), \textrm{Im}(z_2))$ [see Fig.	~\ref{fig1}{(f) and (h)}]. There is a negative region in both the cases, which is a clear evidence of the nonclassical nature of the associated states.
Therefore, the presence of the negative part of the Wigner distribution in almost all the cases illustrates the nonclassical nature of the considered quasi-Werner states. However, it failed to give us a quantitative analysis of the effect of photon addition on the negativity of the Wigner function {in particular and nonclassiality in general}.

\begin{figure}[tbp]
\centering
\includegraphics[width=0.47\textwidth]{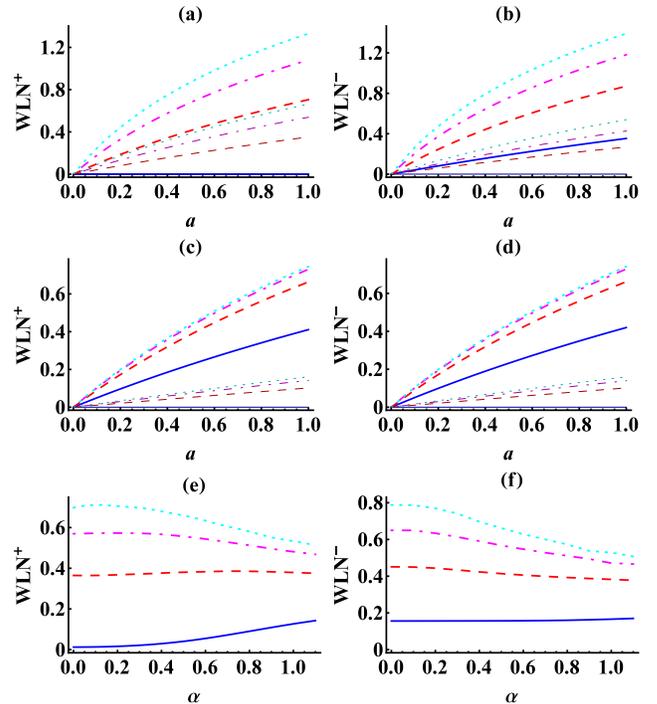} 

\caption{(Color online) In the first (second) column, Wigner logarithmic negativity of the state $\rho(\psi^+, a)$ ($\rho(\psi^-, a)$) is plotted with respect to the mixing parameter $a$ for (a)-(b) $\alpha=0.2$, $\beta=0.1$, (c)-(d)  $\alpha=1.2$, $\beta=1.1$. $\mathrm{WLN^+}$ and $\mathrm{WLN^-}$ are plotted with coherent amplitude $\alpha$ for $\beta=0.67$, $a=0.4$ in (e)-(f), respectively. The smooth (blue), dashed (red), dot-dashed (magenta), and dotted (cyan) lines correspond to $m=0,1,2,3$, respectively. The corresponding thin lines in (a)-(d) represent Wigner logarithmic negativity of the respective reduced single mode states.}
\label{fig2}
\end{figure}

\section{Wigner Logarithmic Negativity}
\label{sec4}

A nonclassicality quantifier based on the negative values of the Wigner function, namely Wigner logarithmic negativity, is defined as \cite{albarelli}
\begin{eqnarray}
\mathrm{WLN}(\rho) = \log\left(\int{|W_{\rho}(z)|\,d^2 z_n}\right), \label{eqWLN}
\end{eqnarray}
where the integration is taken over the whole phase-space $\mathcal{R}^{2n}$ for $n$ number of modes.

The negativity of the Wigner distribution exposes the nonclassical nature of quasi-Werner states in Fig.\,\ref{fig1}, but the quantitative dependence of nonclassicality over different parameters cannot be predicted from that figure. In Fig.~\ref{fig2}, we have shown that the Wigner logarithmic negativity of the photon added quasi-Werner state diminishes with the increasing value of the coherent amplitudes $\alpha$ and $\beta$ [see in Fig. \ref{fig2}{(a)}, $\alpha=0.2$, $\beta=0.1$ while in Fig. \ref{fig2}{(c)}, $\alpha=1.2$, $\beta=1.1$] for $m>0$. However, the opposite is observed for $m=0$. This can be observed from the variation of WLN with coherent amplitude $\alpha$ shown in Figs.~\ref{fig2}{(e)-(f). When more photons are added, an enhancement in the amount of nonclassicality is observed for both the quasi-Werner states, $\rho(\psi^+, a)$ and $\rho(\psi^-, a)$ though this advantage is not significant for larger values of coherent amplitudes [see Fig. \ref{fig2}{(a)-(d)}]. Thus, photon addition is an effective tool for enhancement of nonclassicality of quasi-Werner states formulated with weak coherent states. The thin lines in Figs.~\ref{fig2}{(a)-(d)} represent the WLN of the single mode states corresponding to each subsystem (obtained after tracing out the other subsystem). We can clearly observe that WLN for the subsystems is zero for $m=0$ but increases with the photon addition and {is always} non-zero for all $m>0$. This shows that the photon addition in both the modes of the quasi-Werner states increases local nonclassicality {i.e., the nonclassicality of the single-mode states obtained by tracing out one of the modes of the quasi-Werner state}. Whether photon addition, which is a local operation on a subsystem of two-mode state, can only enhance local nonclassicality or affect quantum correlations in the states as well will be further discussed in the next section. Note that the WLN for two-mode state acts like an upper bound for the respective single-mode WLN in all the cases as the former constitutes of both single-mode nonclassicality as well as correlations. 

\section{Quantum correlations}
\label{sec5}

Here, we discuss three measures of quantum correlations--concurrence, entanglement of formation, and QD--to analyze the effect of photon addition on quantum correlations. 

\subsection{Concurrence}

Concurrence is a widely used measure of entanglement of a composite two-qubit system \cite{woo97}. In general, if $\rho_{XY}$ denotes the density matrix of a bipartite system $XY$, then concurrence is defined as
\begin{eqnarray}\label{eqC}
C(\rho_{XY}) = \mbox{max}\,\,[0, \lambda_1-\lambda_2-\lambda_3-\lambda_4],
\end{eqnarray}
where $\lambda_i\,\,(i=1, 2, 3, 4)$ are non-negative real number and the square roots of the eigenvalues of the non-Hermitian matrix $\rho_{XY}\widetilde{\rho}_{XY}$, arranged in decreasing order. Also, the spin-flipped density matrix is
\begin{eqnarray*}
\widetilde{\rho}_{XY} = (\sigma_y\otimes\sigma_y)\rho^*_{XY}(\sigma_y\otimes\sigma_y),
\end{eqnarray*}
where $\rho^*_{XY}$ is the complex conjugate of $\rho_{XY}$, $\sigma_y$ is the Pauli spin matrix
$
\sigma_y =
\left(\begin{array}{cc}
       0 & -i \\
       i & 0 \\
       \end{array}
       \right).
$

Using Eq.~(\ref{eqC}) the concurrence of the quasi-Bell states (\ref{eqQB}) can be derived as
\begin{eqnarray}
C(\rho^{\pm}) & = & \frac{N_{\pm}^2}{2\,n_+^\alpha n_-^\alpha n_+^\beta n_-^\beta},
\end{eqnarray}
where $\rho^{\pm}=|\psi^\pm\rangle\langle\psi^\pm|$.
Thus, the concurrence for the quasi-Werner state $\rho(\psi^{\pm}, a)$ can be obtained as
\begin{eqnarray}\label{eq:Cpm}
C^{\pm}_a & = & \mbox{max}\,\,\Big[0, \left(\frac{a}{2}\frac{N_{\pm}^2}{n_+^\alpha n_-^\alpha n_+^\beta n_-^\beta}-\frac{1-a}{2}\right)\Big].
\end{eqnarray}

\subsection{Entanglement of Formation}

The entanglement of formation (EOF) is defined as the average entropy of entanglement of its pure state decomposition, minimized over all such possible decompositions. If $\rho_{XY} = \sum\limits_i p_i|\psi_i\rangle\langle\psi_i|$ is the density matrix for a pair of quantum systems $X$ and $Y$ then EOF is defined as
\begin{eqnarray}
E(\rho_{XY}) = \min\sum\limits_i p_i E(|\psi_i\rangle),
\end{eqnarray}
where $E(|\psi_i\rangle)$ is the entropy of entanglement. For an arbitrary two-qubit state, this EOF can be presented as \cite{woo98}
\begin{eqnarray}\label{eqE}
E(\rho_{XY}) & = & H\left(\frac{1+\sqrt{1-C^2(\rho_{XY})}}{2}\right),
\end{eqnarray}
where $C(\rho_{XY})$ is concurrence defined in Eq.~(\ref{eqC}) and binary entropy function
\begin{eqnarray*}
H(x) & = & -x \log_2{x}-(1-x)\log_2(1-x).
\end{eqnarray*}
Thus, EOF for quasi-Werner states can be easily obtained in terms of concurrence reported in Eq.~(\ref{eq:Cpm}).

\subsection{Quantum Discord}
QD \cite{luo08,dhar13} is defined as the difference between two classically equivalent expressions for mutual information after extending to the quantum regime.
If $\rho_X$ and $\rho_Y$ are the marginal states of a bipartite density operator $\rho_{XY}$ shared by parties $X$ and $Y$, the expressions for quantum mutual information are
\begin{eqnarray}
\mathcal{I}(\rho_{XY}) & \equiv & S(\rho_X)+S(\rho_Y)-S(\rho_{XY}),\nonumber\\
\mathcal{J}(\rho_{XY}) & \equiv & S(\rho_X)-S(\rho_{XY}|\rho_Y),
\label{eq16}
\end{eqnarray}
where $S(\rho_{XY})$ is the von Neumann entropy of the quantum state $\rho_{XY}$, and $S(\rho_{XY}|\rho_Y)$ is the quantum conditional entropy. The quantum versions of these two classically equivalent expressions are not equal. To quantify QD, Ollivier and Zurek \cite{olli01} suggested the use of von Neumann type measurements which consist of a set of one-dimensional projectors that sum to the identity operator. Let the projection operators $\left\{\Pi_k^Y\right\}\equiv\{|\pi_0\rangle\langle\pi_0|,\,|\pi_1\rangle\langle\pi_1|\}$ describe a von Neumann measurement for subsystem $Y$ only, where $|\pi_0\rangle = \cos{\theta}|0\rangle+e^{i\phi}\sin{\theta}|1\rangle$ and $|\pi_1\rangle = e^{-i\phi}\sin{\theta}|0\rangle+\cos{\theta}|1\rangle$ with $\langle i|j\rangle = \delta_{ij},\forall\, i,j\in\{0,1\}$. The conditional density operator $\rho_{XY}|\Pi_k^Y$ associated with the measurement result $k$, would then be
\begin{eqnarray*}
\rho_{XY}|\Pi_k^Y = \frac{1}{p_k}(I\otimes \Pi_k^Y)\rho_{XY}(I\otimes \Pi_k^Y),
\end{eqnarray*}
where the probability $p_k=\mbox{tr}[(I\otimes \Pi_k^Y)\rho_{XY}(I\otimes \Pi_k^Y)]$. The quantum conditional entropy with respect to this
measurement is given by \cite{ali10}
\begin{eqnarray}
S(\rho_{XY}|\{\Pi_k^Y\}) \equiv \sum\limits_k p_k S(\rho_k),
\end{eqnarray}
and the quantum mutual information associated with this measurement is given by
\begin{eqnarray}
\label{eq18}
\mathcal{J}(\rho_{XY}) & \equiv & S(\rho_X)-\min\limits_{\{\Pi_k^Y\}}S(\rho_{XY}|\{\Pi_k^Y\}),
\end{eqnarray}
where the minimization is taken over all such possible 1-dimensional projectors $\{\Pi_k^Y\}$.
QD can then be found by using the relations (\ref{eq16}) and (\ref{eq18}) as
\begin{eqnarray}
\label{eq19}
QD(\rho_{XY}) = \mathcal{I}(\rho_{XY})-\mathcal{J}(\rho_{XY}).
\end{eqnarray}
In our case, the quantum states after applying projective measurements over party $Y$ are 
\begin{eqnarray}
\rho_{XY}^{\pm}|\Pi_0^Y
& = & \frac{1}{4p_0^{\pm}}\left( \left[(1-a)+a (\chi_0^{\pm}\cos\theta)^2\right] |0\rangle\langle0|\right.\nonumber\\
& & +\left[(1-a)+a(\chi_1^{\pm}\sin\theta)^2\right] |1\rangle\langle1| \nonumber\\
& &+\left.\left\{a \chi_0^{\pm} \chi_1^{\pm} \cos\theta \sin\theta e^{i\phi}|0\rangle\langle1| +\mathrm{H.c.}\right\}\right)
\end{eqnarray}
and
\begin{eqnarray}
\rho_{XY}^{\pm}|\Pi_1^Y
& = & \frac{1}{4p_1^{\pm}}\left( \left[(1-a)+a(\chi_0^{\pm} \sin\theta)^{2}\right] |0\rangle\langle0|\right.\nonumber\\
& & +\left[(1-a)+a(\chi_1^{\pm}\cos\theta)^{2}\right] |1\rangle\langle1| \nonumber\\
& &+\left.\left\{a \chi_0^{\pm} \chi_1^{\pm} \cos\theta \sin\theta e^{i\phi}|0\rangle\langle1| +\mathrm{H.c.}\right\}\right),
\end{eqnarray}
where $\mathrm{H.c.}$ corresponds to Hermitian conjugate terms, $\chi_0^{\pm}=\frac{N_{\pm}}{n_{\pm}^\alpha n_{\pm}^\beta}$, and $\chi_1^{\pm}=\frac{N_{\pm}}{n_{\mp}^\alpha n_{\mp}^\beta}$. Also, the probabilities are given by
\begin{eqnarray}
\label{eq22}
p_j^{\pm} & = & \frac{1-a}{2}+\frac{a}{4}\left\{(\chi_j^{\pm}\cos\theta)^2+ (\chi_{1-j}^{\pm}\sin\theta)^2\right\}
\end{eqnarray}
for $j\in\{0,1\}$. Thus, the eigenvalues of the matrix $\rho^{\pm}_{XY}|\Pi_j^Y$ are
\begin{eqnarray}
\label{eq23}
\left(\frac{1-a}{4p^{\pm}_j}\right)\quad\mathrm{and} \quad 1-\left(\frac{1-a}{4p^{\pm}_j}\right)\, \forall j\in\{0,1\}.
\end{eqnarray}
Now, using Eq.~(\ref{eq16}) with Eqs.~(\ref{eq19})-(\ref{eq23}), QD for the quasi-Werner state $\rho(\psi^{\pm}, a)$ can be computed as
\begin{eqnarray}
\label{eq24}
QD_a^{\pm}& = & -\left(\frac{1-a}{2}+\frac{a (\chi_0^{\pm})^2}{4}\right)\log_2\left[\frac{1-a}{2}+\frac{a (\chi_0^{\pm})^2}{4}\right] \nonumber\\
& & -\left(\frac{1-a}{2}+\frac{a (\chi_1^{\pm})^2}{4}\right) \log_2\left[\frac{1-a}{2}+\frac{a (\chi_1^{\pm})^2}{4}\right]\nonumber\\
& & +3\left(\frac{1-a}{4}\right)\log_2\left[\frac{1-a}{4}\right]+\left(\frac{1+3a}{4}\right)\nonumber\\
& &\log_2\left[\frac{1+3a}{4}\right] -\sum_j \left\{\left(\frac{1-a}{4}\right)
\log_2\left[\frac{1-a}{4p_j^{\pm}}\right]\right. \nonumber\\
& &-\left.p_j\left(1-\frac{1-a}{4p_j^{\pm}}\right) \log_2\left[1-\frac{1-a}{4p_j^{\pm}}\right]\right\} .
\end{eqnarray}

\begin{figure}[tbp]

\centering

\includegraphics[width=0.47\textwidth]{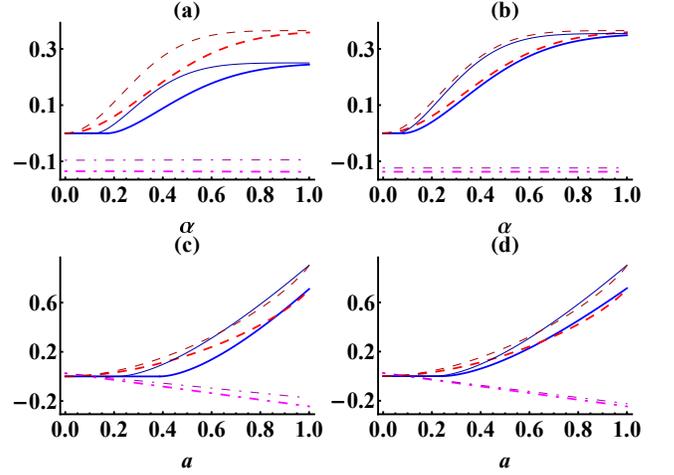} 

\caption{(Color online) Comparison of quantum correlations and Wigner function for the state $\rho(\psi^+, a)$ in Column 1 and $\rho(\psi^-, a)$ in Column 2, for fixed parameter values such as $q_1=0=p_1=q_2$, and $p_2=0.5$. Variation for $\rho(\psi^{\pm}, a)$ with respect $\alpha$ with $a=0.6$ in (a)-(b); and with respect to $a$ with $\alpha=0.5$ in (c)-(d). $\beta=\pi/4p_2$ and $\beta=\pi/2p_2$ for $\rho(\psi^+, a)$ and $\rho(\psi^-, a)$, respectively. The blue solid, red dashed, magenta dot-dashed lines correspond to EOF, QD, and Wigner function without photon addition, respectively. Corresponding thin lines of darker colors show the variation of same quantities after single photon addition.}

\label{fig3}
\end{figure}

Using expressions Eqs.~(\ref{eq:Cpm}), (\ref{eqE}), and (\ref{eq24}), we can quantify the effect of photon addition on quantum correlations in the quasi-Werner states. As EOF is defined in terms of concurrence, it would be sufficient to discuss the former one only to quantify entanglement. In Fig.~\ref{fig3}, we can clearly observe that both EOF and QD increase with increasing $\alpha$ and mixing parameter, which is consistent with the behavior of WLN. {Further,
photon addition has an advantage in the enhancement of quantum correlations [see Fig.~\ref{fig3}]. Thus, photon addition not only increases nonclassicality, it also increases two-mode quantum correlations, namely entanglement and QD.}
Recently {Wigner function is claimed to be useful in revealing quantum entanglement only, but not QD} (\cite{siyouri,taghia} and other papers by the same group of researchers who authored \cite{siyouri}). The present study allows us to verify the role of Wigner function in detecting quantum entanglement. For the sake of argument, we can choose a point in the phase space and study the quantum correlations for the same set of parameters. There is no prescription for the choice of phase space parameters for the Wigner function in \cite{siyouri,taghia}. Minima of the Wigner function in the phase space may be an appropriate choice of parameters. The negative values of Wigner function of quasi-Werner states $\rho(\psi^{\pm}, a)$ can be attributed to the interference terms. Using this fact, the condition for minima of Wigner function can be obtained as $4(p_1\alpha+p_2\beta)=(2j+1)\pi$ and $4(p_1\alpha+p_2\beta)=2j\pi$ for $\rho(\psi^+, a)$ and $\rho(\psi^-, a)$ for real coherent amplitude $\alpha$ and $\beta$ and integer $j$. Notice that for $\alpha=0=m$ the quantum state $\psi^+$ defined in (\ref{eq1}) becomes separable as $N_+|0\rangle\left(|\beta\rangle + |-\beta\rangle\right)$, in that situation, the negativity of Wigner function is a signature of single mode nonclassicality, i.e., superposition of coherent states. Therefore, there is a parametric space corresponding to separable quantum state and non-positive Wigner function.
Assuming $p_1=q_2=0$ we can obtain condition for Wigner minima in terms of relation between $p_2$ and $\beta$. 
Thus, we observed that the Wigner function remains negative while entanglement is zero [cf. Fig.~\ref{fig3} (a)-(b)]. Notice that for $m=0$, WLN for reduced single mode states is zero for state $\rho(\psi^+, a)$, i.e., local nonclassicality is absent [cf. Fig.~\ref{fig2} (a)-(b)]. Thus, Wigner function is not a witness of quantum correlation as claimed in \cite{siyouri,taghia}. The same fact remains valid for $m>0$ [cf. Fig.~\ref{fig3}], where we can observe that the Wigner minima is reducing due to higher-order Laguerre polynomials which creates some craters and peaks at other phase points at the cost of reduction in the Wigner minima.

\section{Teleportation of a single-mode state}
\label{sec6}

To investigate the utility of the photon addition in the quasi-Werner states, we use these states here as quantum resources in a specific quantum information protocol, i.e., quantum teleportation.

\subsection{Teleportation of a coherent state}

In this subsection, we consider the teleportation of a single-mode coherent state by using the quasi-Werner states (\ref{eq_st}) as quantum teleportation channels. The teleportation is conducted and optimized using the Braunstein-Kimble (BK) protocol \cite{kimble}. The success probability of teleporting a pure quantum state is described by the teleportation fidelity ${F} = \mathrm{tr}(\rho_{\mathrm{in}}\,\rho_{\mathrm{out}})$. In the continuous variable quantum teleportation formalism, the teleportation fidelity can be represented by \cite{chizhov}
\begin{eqnarray}
\label{eq25}
{F} = \frac{1}{\pi}\int d^2 \mu\,\,\chi_{\mathrm{in}}(\mu)\chi_{\mathrm{out}}(-\mu),
\end{eqnarray}
where $\chi_{\mathrm{out}}(\mu) = \chi_{\mathrm{in}}(\mu)\chi_{\mathrm{ch}}(\mu^*, \mu)$ with $\chi_{\mathrm{in}}$ and $\chi_{\mathrm{ch}}$ being the characteristic functions of the input state to be teleported and the two-mode quantum channel, respectively. The symmetrically ordered characteristic function of the input coherent state $|\gamma\rangle$ is given by
\begin{eqnarray}
\label{eqCS}
\chi_{\mathrm{in}}(\mu) = e^{-\frac{1}{2}|\mu|^2}e^{\gamma^*\mu-\gamma\mu^*}.
\end{eqnarray}
The performance of a two-mode entangled quantum channel to
successfully transport a single-mode state is measured by computing the teleportation fidelity (\ref{eq25}). The characteristic functions of the two-mode superposed $m$-photon-added coherent states $|\psi^\pm\rangle$ can be calculated as
\begin{eqnarray}
\label{eq26}
\chi_{\mathrm{ch}}^\pm(z_1, z_2)
& = & \mathrm{tr}[\rho_{\mathrm{ch}}D(z_1)D(z_2)]\nonumber\\
& = & \frac{1}{{m!}^2}\frac{e^{|\alpha|^2+|\beta|^2}}{2L_m^\pm(|\alpha|^2, |\beta|^2)}\bigg[\langle\alpha, \beta|a^m b^m D(z_1) D(z_2)\nonumber\\
& & a^{\dagger m} b^{\dagger m}|\alpha, \beta\rangle + \langle -\alpha, -\beta|a^m b^m D(z_1) D(z_2)\nonumber\\
& & a^{\dagger m} b^{\dagger m}|-\alpha, -\beta\rangle \pm \{\langle -\alpha, -\beta|a^m b^m D(z_1)\nonumber\\
& & D(z_2) a^{\dagger m} b^{\dagger m}|\alpha, \beta\rangle +{\textrm{c.c.}}  \}\bigg].
\end{eqnarray}
Using the identity $(a^\dagger-z_1^*)^m = \sum_{p=0}^m{m \choose p}(a^\dagger)^p(-z_1^*)^{m-p}$, we have found
\begin{eqnarray*}
\langle\alpha|a^m D(z_1) a^{\dagger m}|\alpha\rangle
 = e^{-|\alpha|^2-\frac{1}{2}|z_1|^2-z_1^*\alpha}\,m!\,{_1}F_1,
\end{eqnarray*}
where
${_1} F_1$ is the Kummer's hypergeometric function of the first kind \cite{wolfram}. Similarly calculating the other terms in Eq.~(\ref{eq26}) and using $e^x {_1}F_1(a, b, -x) = {_1}F_1(b-a, b, x)$ and $L_n^{(m)}(x) = \frac{(m+n)!}{m!n!}{_1}F_1(-n, m+1, x)$ \cite{kummer}, the characteristic function can be simplified to
\begin{eqnarray}\label{eq:ch}
\chi_{\mathrm{ch}}^\pm(z_1, z_2)
& = & \frac{e^{-\frac{1}{2}|z_1|^2}e^{-\frac{1}{2}|z_2|^2}}{2L_m^\pm(|\alpha|^2, |\beta|^2)}\nonumber\\
& \times& \Big[e^{-\alpha z_1^*}e^{-\beta z_2^*}L_m^\pm(\alpha^* z_1+|\alpha|^2, \beta^* z_2+|\beta|^2)\nonumber\\
& \pm&
e^{\alpha z_1^*}e^{\beta z_2^*}L_m^\pm(\alpha^* z_1-|\alpha|^2, \beta^* z_2-|\beta|^2)\Big].
\end{eqnarray}
Thus, offering the state $\rho(\psi^+, a)$ ($\rho(\psi^-, a)$) as an entangled resource, the teleportation fidelity $F_{\mathrm{coh}}^+$ ($F_{\mathrm{coh}}^-$)) for transmitting a coherent state can be obtained using Eqs.~(\ref{eqCS}) and (\ref{eq:ch}) in Eq.~(\ref{eq25}). The teleportation fidelity for teleportation of coherent state is independent of coherent state parameter $\gamma$ and thus it is same as average fidelity \cite{telF}. 
\begin{figure}[tbp]

\centering
\includegraphics[width=0.47\textwidth]{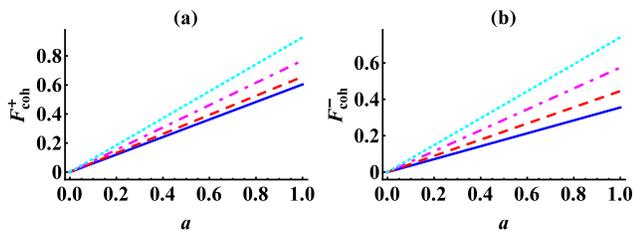}

\caption{(Color online) A comparison of average fidelity of teleporting a coherent state using $\rho(\psi^+, a)$ and $\rho(\psi^-, a)$ for $\alpha=\beta=0.67$ and different photon excitation $m=0$ (blue solid line), $m=1$ (red dashed line), $m=2$ (magenta dot-dashed line), and $m=3$ (cyan dotted line).}

\label{fig4}
\end{figure}

\begin{figure}[tbp]

\centering

\includegraphics[width=0.47\textwidth]{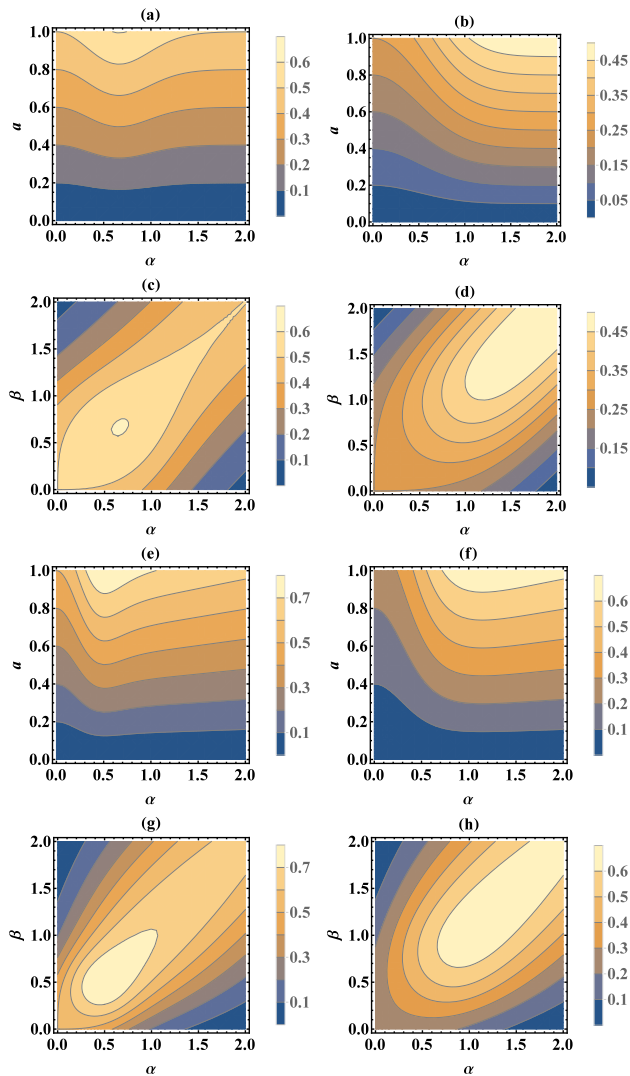} 

\caption{(Color online) Contour plots of teleportation fidelity $F_{\mathrm{coh}}^+$ ($F_{\mathrm{coh}}^-$) in Column 1 (2) as a function of coherent state amplitude $\alpha$ (assuming $\beta$ = $\alpha$) and mixedness criterion $a$ in (a) and (e) ((b) and (f)); with respect to $\alpha$ and $\beta$, for $a=1$ in (c) and (g) ((d) and (h)) for photon excitation $m=0$ and 2, respectively.}

\label{fig5}
\end{figure}

In Figs.~\ref{fig4}{\color{blue}--}\ref{fig5}, we numerically investigate how the average fidelity for teleporting a single-mode coherent state $|\gamma\rangle$ varies with respect to mixing parameter $a$ (ranging from 0 to 1) and coherent state amplitudes $\alpha$ and $\beta$. Figure \ref{fig4} clearly shows that the average fidelity depends on the photon excitation number. The increase in the number of photons added causes the enhancement of the average fidelity in both the cases which is consistent with WLN illustrated in Fig.~\ref{fig2}. For the state $\rho(\psi^+, a)$, the maximum average fidelity is $F_{\mathrm{max}}^+ \approx 0.611$ at $m=0$ while for $m=3$, it enhances to $F_{\mathrm{max}}^+ \approx 0.927$. This clearly indicates that there is an improvement in the maximum average fidelity with increasing $m$. For the state $\rho(\psi^-, a)$, when $m=0$, the maximum average fidelity is $F_{\mathrm{max}}^- \approx 0.361$ which is less than the classical threshold. For larger $m\,(=2)$, $F_{\mathrm{max}}^- \approx 0.565$ crosses the classical bound of teleportation of coherent state 0.5 \cite{telH,himadri}  and thus can be treated as a success for continuous variable quantum teleportation according to the Braunstein-Kimble protocol.  Figure \ref{fig5} shows the variation of average fidelity $F_{\mathrm{coh}}^{\pm}$ with the parameters $\alpha$, $\beta$, and $a$. This verifies that the non-Gaussian photon addition operation improves the fidelity in both the cases. This can be attributed to the entanglement distillation properties of this non-Gaussian operation \cite{ED}. Specifically, the state $\rho(\psi^+, a)$ is preferable over $\rho(\psi^-, a)$ as a quantum channel while transporting a coherent state as it results in higher fidelity. 

\begin{figure}[tbp]

\centering

\includegraphics[width=0.47\textwidth]{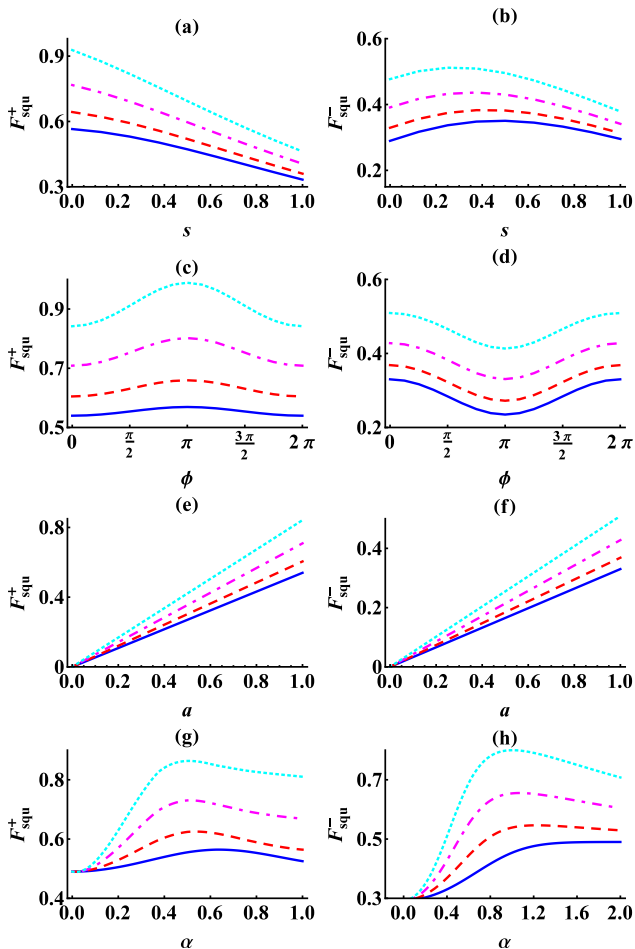}

\caption{(Color online) The dependence of teleportation fidelity $F_{\mathrm{squ}}^+$ and $F_{\mathrm{squ}}^-$ on (a)-(b) squeezing parameter $s$, (c)-(d) phase angle $\phi$ associated with the squeezed state $|\zeta\rangle$, (e)-(f)  mixing parameter $a$ and (g)-(h)  amplitude of the channel parameter $\alpha$ are shown in Column 1 and 2, respectively. Here, we choose $s = 0.2$, $\phi = 0$, $\alpha = \beta = 0.4$, $a=1$, whenever needed. The plots are for different numbers of photon addition as $m=0$ (blue solid line), $m=1$ (red dashed line), $m=2$ (magenta dot-dashed line), and $m=3$ (cyan dotted line).}

\label{fig6}
\end{figure}

\subsection{Teleportation of a squeezed state}

Here we calculate the teleportion fidelity of a single-mode squeezed vacuum state $|\zeta\rangle = S(\zeta)|0\rangle$ with squeezing parameter $\zeta = s\,e^{i\phi}$. The characteristic function for this squeezed state is $\chi_{\mathrm{squ}}(\mu) = e^{-\frac{1}{2} |\mu'|^2}$, where $\mu' = \mu \cosh s+\mu^* e^{-i\phi}\sinh s$. Using a similar approach as above, the fidelity $F_{\mathrm{squ}}^+$ ($F_{\mathrm{squ}}^-$) to send the squeezed state $|\zeta\rangle$ via the entangled channel $\rho(\psi^+, a)$ ($\rho(\psi^-, a)$) is obtained.

Figure \ref{fig6} illustrates the teleportation fidelity for teleporting a single-mode squeezed state via a quasi-Werner state channel with varying parameters $s$, $\phi$, $a$ and $\alpha$. Unlike coherent state,  {where teleportation fidelity was independent of the parameters of the state to be teleported}, the teleporation fidelity  of the squeezed state is found to depend on the parameters of the squeezed state to be teleported, i.e., its squeezing parameter and corresponding phase angle. Thus, average fidelity of teleportation of squeezed state can be obtained by averaging over input state parameters \cite{telF}. We observe that a high fidelity ($\approx 0.91$) can be achieved at the low squeezing regime using $\rho(\psi^+, a)$, which is much larger than that obtained with $\rho(\psi^-, a)$ [cf. Fig.~\ref{fig6}{(a)}-(b)]. The fidelity value drops off for highly squeezed state as the resource requirement for teleporting highly nonclassical state increases. In addition, the phase angle $\phi$ has a limited effect on fidelity. However, $F_{\mathrm{squ}}^+$ is maximum while $F_{\mathrm{squ}}^-$ is minimum for $\phi=\pi$ [see Figs.~\ref{fig6}{(c)}-(d)]. The fidelity increases rapidly with the mixing parameter $a$ and reaches the highest value at $a=1$ as shown in Figs.~\ref{fig6}{(e)}-(f). Figures~\ref{fig6}{(g)}-{(h)} show the variation of teleportation fidelity with parameter $\alpha$ of the quantum channel. The fidelity $F_{\mathrm{squ}}^+$, using $\rho(\psi^+, a)$ as a quantum channel, arrives {at} the approximate maximum value 0.9 at $\alpha \approx 0.4$ and then saturates there. For the state $\rho(\psi^-, a)$, the maximum fidelity value 0.8 is obtained corresponding to $\alpha \approx 0.8$ in photon added quantum channels.

To summarize, it is found that the performance of quasi-Werner state as a quantum channel for teleporting coherent as well as squeezed states is satisfactory. The channel displays high fidelity value, specially when $a > 0.8$ and $m \geq 2$. That means the nonclassical states generated by using superposition of bipartite photon-added coherent states offer a quantum advantage in continuous variable quantum teleportation, also the states provide a better channel for a larger number of photon addition.

\section{Conclusion}\label{sec:Con}
The effect of photon addition on the nonclassical properties of quasi-Werner states defined using superposition of coherent states is studied here with the help of various witnesses of single- anf two-mode nonclassical features, like  Wigner function, WLN, concurrence, EOF and QD. Performance of the quasi-Werner states as quantum channel for teleportation of coherent as well as squeezed states, is also analyzed. The analysis revealed that a local non-Gaussianity inducing operation, namely photon addition, enhances both local nonclassicality (quantified as WLN of the reduced quantum state of a subsystem) as well as quantum correlations (both bipartite quantum entanglement and discord). The WLN for the two-mode state corresponds to the sum of the single-mode WLN (nonclassicality) as well as correlations. Thus, both single-mode nonclassicality and correlations increase due to photon addition. With reference to some of the recent results claiming negative values of the {two-mode} Wigner function as witness of quantum correlations, we observed that for certain choice of parameters the quantum state may be separable but the Wigner function would remain negative. We have given explicit examples to support our argument that two-mode Wigner function may be negative even in absence of quantum correlations due to single-mode nonclassicality in the phase space. 

An enhancement of quantum correlations is shown to have direct consequence in terms of a useful quantum information processing task, continuous variable quantum teleportation. Specifically, teleportation fidelity quantifying the performance of quasi-Werner states as a quantum channel is shown to increase due to photon addition, which may be interpreted as entanglement distillation of the channel due to non-Gaussianity inducing operation. We have observed that $\rho(\psi^+, a)$ is preferable over $\rho(\psi^-, a)$ as a quantum channel for teleportation of a coherent and squeezed state as it results in the higher teleportation fidelity. Further, in view of the recent results \cite{BV1,BV2} reporting stronger correlations present in superposition of coherent states, namely Bell nonlocality, and the present work, we expect that even stronger correlation can be enhanced due to photon addition. We hope the present results will be useful in non-Gaussian state-based continuous variable quantum information processing tasks.  Specifically, we may mention,  implementation of measurement device independent direct communication scheme \cite{MDI} as well as 
teleportation based collective attacks on continuous variable quantum key distribution \cite{anal}, channel purification \cite{pur}, and quantum repeaters \cite{rep}, teleportation can also be used to achieve device independence by circumventing side-channel attacks on the measurement devices in quantum key distribution  \cite{Cry}. Further, it is well-known that continuous variable quantum key distribution would perform better (in the presence of noise) in establishing metropolitan quantum key distribution network. In such an effort, the present analysis is expected to be of use. Keeping that in mind, we conclude the present work with an optimistic view that the present work will lead to a set of interesting results in the context of continuous variable  quantum communication.

\begin{acknowledgments}
KT acknowledges GA \v{C}R (project No.
18-22102S) and support from ERDF/ESF project `Nanotechnologies for Future'
(CZ.02.1.01/0.0/0.0/16\_019/0000754). AP thanks DRDO, India for the support provided through the project number ANURAG/MMG/CARS/2018-19/071.
\end{acknowledgments}

\end{document}